# Anticipating Impacts: Using Large-Scale Scenario Writing to Explore Diverse Implications of Generative AI in the News Environment

Kimon Kieslich[1] , Nicholas Diakopoulos[2] , Natali Helberger[1]

**Abstract**

The tremendous rise of generative AI has reached every part of society – including the news environment. There are many concerns about the individual and societal impact of the increasing use of generative AI, including issues such as disinformation and misinformation, discrimination, and the promotion of social tensions. However, research on anticipating the impact of generative AI is still in its infancy and mostly limited to the views of technology developers and/or researchers. In this paper, we aim to broaden the perspective and capture the expectations of three stakeholder groups (news consumers; technology developers; content creators) about the potential negative impacts of generative AI, as well as mitigation strategies to address these. Methodologically, we apply scenario writing and use participatory foresight in the context of a survey (n=119) to delve into cognitively diverse imaginations of the future. We qualitatively analyze the scenarios using thematic analysis to systematically map potential impacts of generative AI on the news environment, potential mitigation strategies, and the role of stakeholders in causing and mitigating these impacts. In addition, we measure respondents' opinions on a specific mitigation strategy, namely transparency obligations as suggested in Article 52 of the draft EU AI Act. We compare the results across different stakeholder groups and elaborate on different expected impacts across these groups. We conclude by discussing the usefulness of scenario-writing and participatory foresight as a toolbox for generative AI impact assessment.

✉ Kimon Kieslich (corresponding author)
k.kieslich@uva.nl

[1] Institute for Information Law, University of Amsterdam, Amsterdam, The Netherlands

[2] Communication Studies & Computer Science, Northwestern University, Evanston, IL, United States

# 1. INTRODUCTION

Whether overhyped or truly transformative, the growth of generative AI over the last several years has been palpable as it washes over a range of domains from entertainment and law to marketing and news media. The technology leverages a powerful approach to training models from vast quantities of data that can then be prompted to create new pieces of media, whether that's images, text, video, audio, 3D content, and so on, or extract bits of information from input media. The capabilities (and limitations) of generative AI are forcing broad rethinking on how information and media are created and how knowledge work itself is done.

Of particular interest in this work is how generative AI is increasingly used by news organizations and impacts the media environment. News organizations have already started experimenting with it, for example for summarizing content [58], supporting article writing [55], or moderating online content [30,33]. According to a survey conducted among news and media organizations in 2023, 85 percent of the respondents indicated they have experimented with generative AI [6]. The potential for the deployment of generative AI and its use in newsrooms is likely to continue to rise sharply, making discussions of the adverse consequences of large-scale adoption of generative AI, such as job losses, the spread of disinformation through deepfakes [60], accuracy issues (e.g., false source attributions) [22] or an increased offensive use of AI for manipulation or cyberattacks [49], essential. News media companies face the difficult task of navigating the possibilities and limitations of generative AI while maintaining their market position and upholding journalistic quality standards. In light of these potentially detrimental consequences of generative AI use [22,49,60], a systematic assessment of the impacts is a necessary element of any AI strategy. To account for the technological dynamics and societal complexities, such an impact assessment cannot be limited to analyzing the status quo but must be able to anticipate plausible future impacts as well.

The limitations and negative impacts of generative AI aren't only an issue for the *journalistic* field though. Political and legal decision-makers have started to work on regulatory approaches to govern the impacts of AI systems in general, including the use of generative AI in the media sector. Recent regulatory initiatives such as the European Union's draft AI Act or the Digital Services Act adopt a risk-based approach, thereby relying heavily on the ability of policy makers, regulators and regulated parties to be able to anticipate risks to fundamental rights and society [24]. The repeated references to "reasonably foreseeable risks" in the AI Act are a case in point, as are calls by the AI ethics community for more anticipatory studies on AI [17,51].

The responsible development, deployment and governance of AI requires new approaches to prospective research, i.e., research that can anticipate how AI will plausibly develop, what the ethical and societal consequences of this development might be, and how this can be proactively addressed by various stakeholders [12,50,72]. However, anticipating impact is a difficult task as there are an inevitable number of uncertainties that technology development and future prospection bring [13,50,53]. We do not know for certain how technology will develop, how consumers will use it, and what impacts it will have on society in light of potentially complex dynamics and feedback effects. For research in this area, the task is to make estimated projections of plausible future developments that justify AI's implementation in practice [50].

To address these needs, in this paper we develop and refine an approach to study the potential impacts (and their mitigation) of the use of generative AI in the media environment. In particular, we utilize a scenario-writing method in an online survey among EU-member-state residents with a variety of different stakeholders of generative AI technology ($n$=119), reflecting the roles and expertise sets of broad sub-groups of content creators, technology developers, and news consumers to collect cognitively diverse expectations of generative AI's impact on the news environment. Furthermore, we also gauge respondents' ideas about mitigating the outlined harms as well as their opinion on the effectiveness of a specific policy proposal, namely transparency obligations as proposed in the EU AI Act.

In applying qualitative thematic analysis to the scenarios and the additional questions, we demonstrate how a scenario-based method can be a promising approach to conducting impact assessments, particularly due to its ability to create vivid projections and engage a diversity of perspectives. We highlight the alternating perspectives that different stakeholder groups bring in, stressing the need to ensure cognitive diversity in AI impact assessment. Further, we not only systematically map potential negative impacts, but also leverage the diversely sampled survey responses to help illuminate mitigation strategies to counter adverse consequences of generative AI. As such, our study contributes to the scientific literature on assessing AI impact, but also contains practical implications for policy makers or technology developers who are tasked with mitigating harms of such systems.

## 2. RELATED WORK

In this work, we address the aforementioned need for approaches to anticipate AI impacts. In the following subsections we examine the related work on anticipatory governance which motivates our study, the range of existing AI impact assessment approaches, the need for participatory approaches, and the background on the use of scenario-writing as the specific approach we leverage.

## 2.1. Anticipatory Governance

The anticipation of the impact of technologies has been studied under the theoretical approach of anticipatory governance, which examines social impacts of technology in the early development phase [21,29,34]. One aim of anticipatory governance approaches is to help mitigate the uncertainty that inevitably surrounds emerging technologies [13]. In doing so, anticipatory approaches can illuminate the positive and negative aspects about the technology at hand [13] and are as such deeply connected to normative values of how society wants to deal with the respective technology [50]. Anticipatory approaches try to come to a point where technology development acknowledges potential risks and tries to mitigate these beforehand [34,63]. Furthermore, anticipatory governance recognizes that the future can not be *predicted* per se and, thus, aims for showing *possible* future scenarios [56].

Methodologically, anticipatory governance is concerned with navigating the choice between different policy options and enabling the formation of a joint vision of how society wants to engage with technology [34]. Practically, anticipating the impact of emerging technology should be initiated where a technology "is sufficiently developed for meaningful discourse to be possible about the nature of the technology and its initial uses, but where there is still uncertainty about its future implications" [50]. In the deliberation process it is not about discussing certain future events, but to discuss how future scenarios align with (public) values. This depicts a translation process, in which underlying values in the form of reactions to exemplary scenarios are discussed that, in turn, are symbolic for plausible futures and can subsequently inform governance processes [50]. In making socio-technical consequences salient and agreeing on shared values for how the future ought to be, anticipatory governance enables empirical and value-oriented decision-making for stakeholders [34,50].

## 2.2. AI Impact Assessment

In a similar vein, scholars in the field of AI have proposed AI impact assessment to identify and anticipate the impact of AI technology [3,52]. Or, in the words of Selbst [63]: "An Algorithmic Impact Assessment is a process in which the developer of an algorithmic system aims to anticipate, test, and investigate potential harms of the system before implementation; document those findings; and then either publicize them or report them to a regulator". In recent years numerous proposals on how to assess AI impact have been published (for an overview, see [68]). Impact assessments are particularly needed for novel technologies, where societal consequences have yet to be determined or are generally hard to measure [63]. That also means that impact assessments aim to identify risks that are not purely technical and are rather related to the sociotechnical nature of AI [52]. In conducting impact assessments, the approach can potentially detect "errors that would otherwise arise at unpredictable times and characterize performance in the long-tail of errors that is currently opaque" [3] (p. 134). Impact assessments refer to impacts that might affect individuals or groups of society [48,52]. In understanding AI as a sociotechnical system, impact classification should also not be limited to the technical components of AI systems, but should include the interplay between humans or society, respectively, and machines.

Existing AI impact assessment studies apply various methods to identify and categorize these impacts. One way to approach this is by conducting a literature review on existing studies focusing on AI impacts. Shelby et al. [65] performed a scoping review of the computer science literature to identify five groups of AI harms (representational, allocation, quality of service, interpersonal, social system). Hoffman & Frase [37], who distinguish between tangible and intangible harms, issues and events, and AI and non-AI-harms, develop their impact framework based on a review of AI incident reports and subsequent discussions with stakeholder organizations. Explicitly analyzing LLMs, Weidinger et al. [71] performed a multidisciplinary literature research of scientific papers, civil society reports and newspaper articles, to identify 21 risks that can be grouped into six categories including discrimination, exclusion and toxicity (e.g., representational harm), information hazards (e.g., data leaks), misinformation harms (e.g., disseminating misinformation), malicious uses (e.g., users actively engaging in harmful activity), human-computer interaction harms (e.g., manipulation), and automation, access, and environmental harms (e.g., environmental costs). In regard to text-to-image (TTI) technology (e.g., Dall-E, Midjourney), Bird et al. [8], relying on a literature review, distinguish between three broad risk categories (discrimination and exclusion, harmful misuse, misinformation and disinformation) and six stakeholder groups (system developers, data sources, data subjects, users, affected parties, and regulators) that build a framework for analyzing harms of TTI use.

Another way to apply impact assessment is to let users decide on the scenarios and impacts they want to focus on and provide them with a tool that helps them in mapping potential impacts. The AHA! (Anticipating Harms of AI) framework of Buçinca et al. [14] serves as a toolbox to create negative AI impact scenarios for different stakeholder groups. It requires the input of a user (e.g., developer) to first describe a deployment scenario and definition of potential risks to then systematically map out potential stakeholders and concrete impacts.

Another approach commonly taken in AI impact assessment is to collect the opinions of experts, sometimes also referred to as Delphi method. For instance, Solaiman et al. [67] conducted workshops with experts from different backgrounds, including researchers, government and civil society stakeholders as well as industry experts. As a result, they list trustworthiness and autonomy, inequality, marginalization and violence, concentration of authority, labor and creativity, as well as ecosystem and environment as potential evaluation categories for generative AI's impact on society and people.

However, all of these outlined approaches are top-down in the sense that they rely on domain experts or scientific viewpoints, an issue that has been acknowledged in potentially limiting the perspectives available [10]. What is currently missing are approaches that engage individual members of society and invite them to reflect on impacts without presenting them with a potential impact framework beforehand. In our study, using scenario-writing in a participatory foresight approach, we therefore pursue a bottom-up approach to help enrich the field of AI impact assessments.

## 2.3. Participatory Foresight

In considering an anticipatory governance approach and in light of the prior work on impact assessment it is crucial to ask: *whose* prospections should guide anticipatory governance? An approach to leverage cognitive diversity in developing prospections is *participatory foresight* [12,54]. The idea of participatory foresight is to engage a diverse set of participants in anticipating future impact as studies have shown that expert foresight is prone to be biased [5,10,12]. Or, in the words of Metcalf and colleagues [48]: "The tools developed to identify and evaluate impacts will shape what harms are detected. Like all research questions, what is uncovered is a function of what is asked, and what is asked is a function of who is doing the work of asking." (p. 743) By establishing a deliberative social dialogue, a discussion about desirable futures can evolve that depict alternative visions of the future [54]. This also includes the perspectives of laypersons as they "bring the specialists' knowledge down to earth and foresee its possible side-effects in everyday life. Thus they would illustrate the whole complexity of the social and cultural consequences, caused by the triumph of advanced techno-knowledge" [54]. In this way potential blind spots can be filled and pluralistic visions of the future can emerge that reflect the realities and circumstances of all involved – and not only one dominating group. Thus, participatory approaches can be seen as a way to democratize AI impact assessment [62].

Ultimately participatory foresight follows the goal to establish pluralistic perspectives from cognitively diverse individuals with different backgrounds, experiences, and expertise. Consequently, this will result in a far more nuanced and relatable picture of future AI impact. We stress that impacts of generative AI are not only defined in terms of technological issues, but also societal consequences like social inequalities, and human and infrastructural social impacts [67].

But how can society be included in anticipatory studies? In this work we develop an approach based on scenario-writing methods, which is outlined in the next section.

**2.4. Scenario-Writing**

Scenario-writing can be defined as a method of anticipatory thinking, which can be used to outline a variety of different futures [2,3,11,15,64]. Scenarios describe in a creative way (e.g., through stories or motion pictures) future developments that are based on plausible and logical developments [2,11,32]. Importantly scenario-writing enables practitioners and researchers to envision a broad scope of different future alternatives [2,61]. As such, scenario-writing explicitly tries not to *predict* future events (as it is deemed as not possible), but acknowledges the uncertain character of the future and instead focuses on *plausibility* [57]. Further, scenarios can encompass a holistic, sociotechnical view that illuminates the relationship between different actors. Scenarios are placed in a real-word environment [2] and help "to reduce the overabundance of available knowledge to the most critical elements, and then blend combinations of those elements to create possible futures." [15] (p. 49). Moreover, scenarios are especially suitable for identifying novel issues and impacts [2].

Börjeson et al. [11] distinguish between three categories of scenarios namely, *predictive (what will happen?)*, *explorative (what can happen?)* and *normative (how can a specific target be reached?)* scenarios. In regard to anticipatory governance and AI impact assessment, *explorative* scenarios are most suitable for extrapolating future prospections from a starting point in the present and can serve as a "framework for the development and assessment of policies and strategies" [11] (p. 727). Burnam-Fink [15] highlights that narratives are a common technique to make scenarios accessible to a broader audience. Thereby, 'good' scenarios, i.e. scenarios that are trusted and taken seriously by the audience (e.g., decision-makers), are characterized through a plausible story with compelling characters that makes future developments easy to understand [64]. This method has shown to be fruitful, for example as shown by Diakopoulos and Johnson, who used scenario-writing to explore the potential impact of deepfakes on the US presidential elections in 2020 [21].

Moreover, Meßmer and Degeling [47], in the use case of auditing recommender systems, discuss scenario definition as a crucial step to anticipate systemic risks.

## 3. METHOD
### 3.1. Procedure & Measurement

We conducted an online survey with an integrated scenario writing exercise in which we recruited targeted sub-samples of EU residents, content creators, and technology developers (total $N$=156; $n$=52 participants for each group) living in member states of the European Union (EU). Crowdsourcing has proven to be a valuable method to identify diverse social impacts of AI systems [5], and the targeted sub-samples are intended to draw in different types of expertise and experience. Scenario-writing is an established method as well, which is used to capture prospections about the future [21] and can be implemented in surveys [11].

After receiving institutional ethics approval for our study, we created an online survey using the survey tool Qualtrics to assess the scenarios. The questionnaire was constructed in English language and structured as follows: First, respondents were introduced to the study's objective and were informed about data collection and storage. After giving their informed consent to participate in the survey, respondents had to indicate some demographic information (gender, age, educational level, ethnicity, employment sector) as well as some information about AI related attitudes. Following this, participants were introduced to the scenario writing exercise. Figure 1 shows the information we presented. Note that the survey instructions were the same for each respondent group, except for the main characters that respondents were asked to imagine. Respondents were tasked to write from the perspective of news consumers, and content creators as well as technology developers out of their respective expertise perspective. Following that, we offered participants some additional information on generative AI, outlining some of its capabilities (tasks generative AI can fulfill; media formats; examples), limitations (accuracy; attribution; biases) and trends. The full description of the technological information we give to the respondents can be found in the Appendix.

In a last introduction step to the task, we provided respondents with information about the evaluation criteria of the scenarios. We highlighted that scenarios should be creative, specific, believable, and plausible as suggested by Diakopoulos and Johnson [21]. In addition to their base pay for doing the survey we offered a bonus payment of 2£ for the top 10% of scenarios as assessed by the authors according to these quality criteria. We tasked respondents to confirm that they understood the information presented to them and measured the time that respondents engaged with the introductory material.

---

In this exercise we ask you to write a short (~ 300 word) fictional **scenario** to explore how **generative AI** technology could create **risks or negative impacts** for the **news environment** in **five years** from now.

A scenario is a **short story** that includes: (1) a **setting** of time and place, (2) **characters** with particular motivations and goals, and (3) a **plot** that includes character actions and **events** that lead to some impact of interest.

**Generative AI** refers to a technology that can create new content (e.g. text, images, audio, video) based on the content it was trained on.

By **news environment** we mean all of the things related to how news information is produced, distributed, and consumed in society.

Specific Instructions:
- Please develop your scenario **based on your own personal and/or professional perspectives, experiences, and knowledge.**
- Please choose your main **character** to be a [**news consumer | content creator | technology developer**]. You may also consider how the actions of different characters are related to each other. Some other actors in the news environment to consider include: journalists, news organizations, advertisers, technology developers, and platforms for social media and search.
- Please choose your **setting** to be five years in the future and in the society where you currently reside.

---

Figure 1: Task Description

On the following survey page, respondents typed in their scenario. To make the instructions clear, we repeated the instructions shown in Figure 1 on the page. Additionally, respondents had the opportunity to click the back button and read the information about the technology and the evaluation criteria as well. To discourage people from using generative AI tools to fulfill the task for them (which is a growing issue for certain kinds of online tasks [70]), we disabled copying text into the open text field such that people needed to type the story directly into the text box. Additionally, to secure adequate length of the scenarios, we set a minimum character number of 1,000. Participants could choose to write their scenario in English or their native language.

To gather respondents' thoughts about impact mitigation, we asked participants the following after submitting their scenarios: *What could be done to mitigate the risks outlined in the scenario? Please also indicate who you think should be responsible for mitigating the harm. This could be the characters in your story, but it could also be other people/organizations. Please write at least 50 words.*

Additionally, we gathered respondents' ideas about the potential influence of a policy approach that could be introduced to mitigate negative impact, namely transparency obligations. We based this mitigation strategy on Article 52 of the draft EU AI Act [26]. We asked: *Organization/persons that use generative AI systems have to fulfill the following legal obligation: Transparency/Disclosure: The use of generative AI must be made transparent, i.e. it must be clearly and visibly disclosed that generative AI was used in the creation of content. This can be done, for example through labeling. With the information provided, would the transparency requirement change your scenario? Please write at least 50 words.*

Lastly, respondents were asked to evaluate the scenario writing task in regard to difficulties in the writing process and comprehensiveness of the introductory material. Consequently, respondents were thanked and debriefed. We paid 11£ for participation in the study based on an estimate for the expected time for completion of the task and a reasonable wage[1].

### 3.2. Pre-Test

To pilot our method and get feedback on the scenario task we conducted two in-person workshops with more than 30 participants. Based on workshop feedback we fine-tuned our research instrument in terms of task clarity as well as adjusting the content and the framing of the information presented. We then set up the online survey.

We pre-tested the online survey with 30 participants (news consumer sample) utilizing the survey panel provider Prolific. Data collection was conducted on July 21, 2023 with participants from one EU country (the Netherlands) and with a balanced sex distribution. Based on a close reading of the scenarios as well as the feedback of the respondents, we substantially revised the instructions for the scenarios. In particular, we decided to specify the role of the main character for each group (news consumer, content creator, technology developer) as respondents indicated having trouble imaging characters that they were not familiar with. Moreover, we restructured the presentation of the introductory material to make it more accessible. The changes resulted in the task as presented in Figure 1.

---

[1] We based our remuneration on the minimum wage in the Netherlands where the pre-test was conducted.

**3.3. Study Sample**

In deploying the refined version of our survey, we collected 156 scenarios via Prolific. We collected 52 scenarios for each stakeholder group in the time span of August 3, 2023 to September 10, 2023. For news consumers, we chose to sample for EU residents as a large majority of people in the EU consume news information[2] and even those who do not formally consume it are still exposed to related information in the media environment . Content creators and technology developers were specifically selected using Prolific's sampling criteria. Content creators were defined as people that indicated at least one of their (current or former) *employment roles* as *journalist*, *copywriter/marketing/communications role*, and/or *creative writing*. For the survey of technology developers, we used the *employment sector* "*information technology*" provided by the survey platform as a selection criterion.

We decided to survey residents of all EU states since the EU, with the EU AI Act [24], on the one hand offers a joint approach to the development and use of AI, and, on the other hand, still leaves room for diversity given the cultural, political and socio-economical differences of the EU member states. Thus, we aimed for recruiting a variety of people living in a broad range of EU member states. To ensure diversity of EU member states, we grouped all EU countries into three groups based on the annual average salaries in the respective countries [27] and surveyed 17 respondents[3] in each group in each stakeholder group[4]. We chose this indicator because a soft launch (conducting the survey with a small subset of respondents to test its functioning) showed that a disproportionate number of respondents from low-income countries participated in the survey. This is plausible given that we decided to pay a fixed remuneration for taking part in the survey despite income differences in the EU member states that make that remuneration comparatively more or less valuable.

---

[2] https://ec.europa.eu/eurostat/web/products-eurostat-news/-/ddn-20220824-1
[3] We approved three additional scenarios from respondents that, due to a technical error, were not counted as completed by Prolific. We paid the respective respondents and added the scenarios to the corpus, thus, resulting in a sample of 156 scenarios.
[4] Low income group: Bulgaria, Croatia, Czech Republic, Greece, Hungary, Latvia, Poland, Romania, Slovakia; Medium income group: Cyprus, Estonia, Italy, Lithuania, Malta, Portugal, Slovenia, Spain; High income group: Austria, Belgium, Denmark, Finland, France, Germany, Ireland, Luxembourg, Netherlands, Sweden

To promote gender diversity, we used Prolific's built-in feature to help balance the gender distribution of the sample for each sub-group.[5]

Further, we only invited those respondents to participate in the survey, who indicated they speak English fluently. As in the pretest, the survey was conducted in English. However, participants had the opportunity to write their scenarios in their native language as we expected a higher quality of scenarios if participants felt comfortable with the language. We translated all scenarios using DeepL, a machine translation product with high performance. Automated translations have recently been established as a viable method in communication research [19,45] and we reason here that any minor translation issues would not create a threat to the validity of our subsequent thematic analysis.

### 3.4. Data Filtering

For data cleaning, first we identified scenarios that were thematically out of the scope of the task, i.e. that did not refer to the news environment or generative AI technology. Following this procedure, we removed twelve scenarios for the news consumer group, four from the content creator group and four from the technology developer group.

Second, although we took steps to prevent the use of LLMs in the scenario writing exercise, we can not completely rule out the possibility of respondents using LLMs. Thus, we decided to flag scenarios with four criteria that could indicate such use. We filtered out those cases that were flagged from at least two of the criteria. First, we checked all scenarios for their likelihood of being written by an AI tool with GPTZero [69][6] and flagged scenarios with a likelihood of being AI-generated of 50 percent or higher. Second, we flagged all scenarios that were written in under 20 minutes as this was determined in our in-person workshops as the minimum amount of time respondents needed to compose a reasonable scenario.[7] Third, we flagged those scenarios in which respondents read the instructions in under two minutes time. Again, the time flag was based on our experiences at the pre-test workshops. Fourth, we manually flagged scenarios as potentially being written by ChatGPT after reading them in detail. Individual flags are based on our own experiences prompting ChatGPT to write scenarios and the emerging patterns produced by those prompts. For example, if prompted with the instructions for our task ChatGPT often generates a very similar structure of output text. After applying these four flags and filtering those scenarios out with two or more flags, this resulted in an elimination of 17 scenarios, leaving us with 119 scenarios as the final sample for analysis.

---

[5] Prolific only offers "sex" as a filter variable for equal distributions among subsamples, using the response to the question: 'What is your sex, as recorded on legal/official documents?' Participants answer this question with one of two options: [Male / Female]' https://researcher-help.prolific.com/hc/en-gb/articles/360009221213-How-do-I-balance-my-sample-within-demographics. Although Prolific's system only operates according to binary sex, in order to ensure representation of non-binary people, we included a more inclusive query for respondents' gender, including the option "non-binary", "prefer to self-describe" and "prefer not to answer". We used this measure to describe our sample (see Table 1).

[6] GPTZero is mostly trained on English language text and, thus, is less accurate for other languages. However, we assume that, if respondents decided to use LLMs to write their scenarios, the output text would most likely be in English as a) copying the task description and running it provides an English output text and b) even if they provide a task description before inserting our task description still provides an English output. Thus, participants would need to either specify that they want the LLM output in their native language or translate our task description before inserting it in a prompt. We believe that both ways are too time-intensive, especially for respondents that aim to reduce workload by using LLMs.

Table 1: Sample Description

|  | News Consumer (n=35) | Content Creator (n=41) | Technology Developer (n=43) |
|---|---|---|---|
| Gender | women=16<br>men=17<br>non-binary=2 | women=19<br>men=22<br>non-binary=0 | women=20<br>men=22<br>non-binary=1 |
| Average Age | 30 | 34 | 35 |
| Educational level | Bachelor degree or higher=21 | Bachelor degree or higher=30 | Bachelor degree or higher=33 |
| Race | White=29<br>Mixed=5<br>Other=1 | White=40<br>Black=1 | White=37<br>Mixed=2<br>Asian=2<br>Other=2 |
| Country of residence | Portugal=10<br>Poland=8<br>Netherlands=3<br>France, Germany, Greece, Italy=2<br>Denmark, Finland. Hungary, Ireland, Spain, Sweden=1 | Poland=12<br>Italy, Netherlands, Spain=4<br>France=3<br>Belgium, Finland, Germany, Greece, Portugal=2<br>Hungary, Luxembourg, Slovenia, Sweden=1 | Portugal=8<br>Poland, Germany=6<br>Italy, France=4<br>Greece, Hungary=3<br>Belgium, Czech Republic, Ireland, Spain=2<br>Austria=1 |

---

[7] We also applied a statistical approach in determining outliers. However, given the skew in the data (many high outliers above the average time score), applying a statistical approach like mean-2SD resulted in negative time values and, thus, was not possible for our study.

Table 1 describes the sample statistics for each sub-sample. The samples predominantly consisted of white and well-educated respondents, a limitation we will return to in our discussion. Furthermore, men are slightly overrepresented in the content creator and technology developer sample. In addition, we managed to sample for a broad range of different countries.

**3.5. Analysis Methods**

We analyzed the final set of scenarios using a qualitative thematic analysis approach including open and axial coding of themes. We constantly and iteratively used open coding techniques to gather excerpts of the scenarios, structure and typologize these and applied constant comparison to reevaluate the emerged codes and discuss them amongst the authors [20,31,46]. We additionally wrote memos to structure the interpretations of the findings.

In detail, the lead and second author of this paper read and coded 30 scenarios (10 for each stakeholder group) independently and derived the first set of open codes and impact classifications. Then, both authors compared and discussed their classification scheme and created an adapted and joint version of the classification scheme, which included the codes of both authors as well as the dimensions of "impact scope" and "agency". Afterwards, the first author coded all remaining scenarios and enlarged the impact categorization scheme for new, emerging codes (impact themes and specific impacts). All corresponding quotes that were used for identifying a code were collected in a structured document, containing all impact themes and specific impacts. After coding all scenarios, the author team discussed and edited the categorization scheme. All authors read all excerpts and quotes derived from the scenarios and discussed the impact classification scheme and governance strategies classification scheme. After three extensive, consecutive meetings, agreement on the classification schemes were reached.

**4. FINDINGS**

We structure our analysis along two themes. First, we will elaborate on the *impacts* enumerated in the scenarios. Second, we will outline *mitigation strategies* that were addressed by the respondents. In addition, we will also discuss the effectiveness of a specific mitigation strategy to counter the negative impacts of the scenarios: transparency obligations. In each section, we will compare the codes of the different stakeholder groups and highlight similarities as well as differences.

**4.1. Enumerating Impacts**

The scenarios express a multitude of different impacts. Table 2 lists the various impact themes and the specific impacts in each that we observed, together with illustrative examples. In the following, we will elaborate on each of the impact themes. We also highlight differences and similarities between the stakeholder groups (i.e. news consumers, content creators, and technology developers) in how they raise awareness of the specific impacts.

In addition, we identified two *dimensions* that help to capture variance applied to different impact themes and are used to elaborate those descriptions below. First, we observed the *impact scope* of each specific impact. We define impact scope at three levels. Impacts can either occur at an individual (e.g., main characters), an organization (e.g., newsrooms), or a societal (e.g., political system) level. Where applicable, we discuss which impact scopes are (mainly) addressed within each theme. Further, we also identified *agency* as another dimension of impacts. We distinguish between *causal* agency, *intentional* agency, and *triadic* agency as proposed by Johnson & Verdicchio [38]. In their terminology, causal agency can be ascribed to AI systems as they can cause impacts, i.e. shape the world. However, AI systems, as they are technological artifacts, can not have intentions. Thus, the dimension of intentional agency can only be ascribed to humans. Intentional agency includes causal agency, but adds a layer that refers to a mental state of humans. As such intentional behavior can also cause impacts. Though, in terms of responsibility in an ethical sense, intentionality can be traced back on various mental states. For instance, stakeholders can act out of a malicious intent, but can also be negligent in a sense that they fail to anticipate negative impacts. They can also act in well-intentioned ways, but cause negative impacts. Intentional agency of human stakeholder can also be combined with causal agency of AI systems. This interplay is called triadic agency. Specifically, Johnson and Veridcchio describe it as follows: "Humans and artifacts work together, with humans contributing both intentionality and causal efficacy and artifacts supplying additional causal efficacy."[38] In the following, if applicable to the impact themes, we will outline how agency plays a role in causing the impacts identified. We note that agency of characters can also be well intentioned and relating to mitigation strategies that are discussed in section 4.2.

**Table 2: Impacts**

| Impact Themes | Specific Impacts | Example Quotes<br>NC=News Consumer; CC=Content Creator; TD=Technology Developers; S=Scenario Number |
|---|---|---|
| Well-Being | *Addiction*$^{NC,CC}$; *Reputation*$^{NC,CC}$; *Physical Harm*$^{NC,TD}$; *Mental Harm*$^{NC}$ | "Artur is plunged into a nightmare. The outrage is fast and furious. Despite his protestations of innocence, the evidence seems insurmountable." (TD S25) |
| Labor | *Competition*$^{NC,CC}$; *Job Loss*$^{NC,CC,TD}$; *Loss of Revenue*$^{NC,CC}$; *Changing Job Roles*$^{NC}$: *Unemployment*$^{NC,TD}$, | "Marcos eventually realized that he cannot compete with AI.[...]. For a long time, he resisted and thought he could go on working as usual, but nobody called him anymore. He was an old and inefficient man who could no longer do much." (CC S2)<br>"All the journalists had to learn these new technologies and those who did not quickly became disposable and were fired." (TD S41)<br>"Traditional news organizations struggled to regain their footing in this environment, their commitment to fact-checking overshadowed by the speed of AI-generated news." (CC S13)<br>"Looking at his coworkers he sees there's a stagnation on his floor, people are changing the focus from delivering solid news, to being the first to release news. Any news. Being the fastest and first to publish has become the most important thing." (CC S50) |
| Autonomy | *Loss of Orientation*$^{NC}$; *Loss of Control*$^{TD}$; *Machine Autonomy*$^{NC}$ | "There are so many different opinions and streams of information coming from so many different GenAIs that he doesn't know which way to turn." (NC S14) |

| Legal Rights | *Copyright Issues*[NC]; *Legal Actions*[CC,TD], *Freedom of Expression*[TD], *Lack of Regulation*[TD] | "'That's my drawing'," he thought in surprise and remembered which drawing specifically he was referring to. [...] Adam had no doubt that his illustration had been used and processed without his permission." (NC S43) <br> "This practice has faced criticism in recent years, but with use of AI rapidly expanding and legislation severely lagging behind on curbing it, disinformation in the media is at an all-time high." (CC S1) |
|---|---|---|
| Media Quality | *Accuracy / Errors*[NC,CC,TD]; *Loss of Human Touch*[CC,TD]; *Sensationalism*[NC,CC,TD]; *Credibility / Authenticity*[CC,TD]; *Lack of Diversity / Bias*[NC,CC,TD]; *Clickbait*[CC,TD]; *Journalistic Integrity*[TD]; *Reframing*[CC,TD]; *Attribution*[NC,CC]; *Distinction between Journalism & Ads*[CC]; *Lack of Fact-Checking*[CC,TD]; *Explainability*[CC]; *Superficiality*[NC]; *Over-Personalization*[CC,TD]; *Ethics*[NC]; *Accountability*[NC,TD] | "The quest for speed and quantity started overshadowing the need for accuracy and quality." (NC S20) <br> "The AI words were shallow, often overlooking the finer points a human reporter would grasp." (CC S48) <br> "The media is in turmoil as journalists struggle to uncover the truth in a deluge of sensationalism." (TD S25) <br> "The AI's algorithms lacked ethical considerations and inadvertently amplified biased narratives, contributing to the polarization of society." (CC S13) <br> "Deviations from the truth no longer existed in principle. Anyone could create any narrative they wanted and AI authenticated it all, made it real, whatever the facts were." (CC S22) <br> "As human journalists struggled to compete, diverse perspectives and in-depth investigations became rare commodities." (CC S13) |
| Security | *Hacking*[TD]; *Cybersecurity*[NC] | "It turns out that the radio station had been hacked by a cybercriminal group related to former journalists, news anchors and producers." (NC S21) |
| Trustworthiness | *Discern of Fact and Fiction*[NC,CC,TD]; *Media Fatigue*[NC]; *Mistrust*[NC,TD]; *Over-Reliance on AI*[NC,TD]; *Information* | "But how much of this was real? How much of it was altered by AI or the people maintaining it? How much was exaggerated by the AI for dramatic effect? How much of it was downplayed not to cause too much concern among the viewers?" |

| | *Chaos*[CC] | (NC S49)<br>"The public trust in the media has been utterly shattered - for a good reason, too." (TD S3660dc) |
|---|---|---|
| Political | *Fakes News / Misinformation*[NC,CC,TD]; *Political Consequences*[NC,CC,TD]; *Opinion Monopoly*[NC]; *Manipulation*[NC,TD] | "The use of AI to create deep fakes of politicians, videos to doctor events that never even occurred and the orchestrated spread of disinformation through various foreign media channels is starting to seriously challenge the status quo of several western democracies." (CC S45)<br>"The news spread like wildfire impacting global markets and causing unnecessary panic. This incident revealed that AI's speed compromised the verification process, leading to misinformation outbreaks." (NC S20)<br>"He got up and looked out the window of his house thinking: what world was this where people can manipulate videos to this extent?" (NC S12) |
| Social Cohesion | *Polarization*[NC,CC,TD]; *Dissatisfaction*[CC]; *Social Divide*[NC,TD]; *Discrimination*[NC,TD]; *Real-World Conflicts*[CC,TD] | "Friends and family are increasingly divided, and heated debates ensue, fueled by the contrasting perspectives presented by InfoSift. People become entrenched in their beliefs, as the algorithm continually reinforces their existing viewpoints, creating echo chambers." (NC S33)<br>"The tensions started to mount and mount until the first fights broke out." (NC S18) |
| Education | *Literacy*[NC,CC,TD]; *Critical Engagement*[NC] | "He knew that the vast majority of readers were not educated enough to realize that; he knew that way too many would believe anything they would read. He knew swathes of people would not question whether an image was real. If it's on the news outlets, it must be real, right? "(TD S3) |

Note: Superscripts denote the occurrence of the code in the corresponding stakeholder group: NC=News Consumer; CC=Content Creator; TD=Technology Developer

*4.1.1. Well-Being*

The scenarios described four different forms of impact on individual well-being. The mentioned mental impacts reached from negative emotions to severe mental illnesses (e.g., depression) that were caused by generative AI, for example, through online harassment based on fake news/images. Strongly connected to mental harm is also the addiction (e.g., to social media apps powered by generative AI) that some scenarios described. Reputational damage, for example, caused by a fake news campaign against a politician/journalist, but also a (news) organization, was also outlined in some scenarios. Reputational damage, then, was also oftentimes connected to mental harm. Lastly, some scenarios even pointed out physical harm (e.g., suicide) that was based on mental harm issues caused by generative AI. All in all, well-being impacts were prevalent among the news consumer scenarios, whereas content creators and technology developers only addressed those impacts at a marginal level.

Well-being impacts mostly occur on an individual level as they are related to personal consequences for characters. However, some impacts also address the organizational level, for instance, if the reputation of a news company is damaged as a result of generative AI use.

*4.1.2. Labor*

Labor impacts are addressed frequently in the scenarios of all stakeholder groups. We found five sub-codes within the labor impact theme, namely *competition*, *job loss*, *unemployment*, *loss of revenue*, and *changing job roles*.

Some scenarios mention stronger competition due to the introduction of generative AI in the newsroom. Generative AI is expected to be a competitor as it replaces more and more tasks that were previously performed by humans. Competition, then, is highly interrelated with the notion of changing job roles. Some scenario-writers elaborate on economic pressure for content creators to learn new skills in order to adjust to the changes in the media environment. On the other hand, strong competition is also connected to the more severe impact of job loss that was one of the most prevalent codes among all scenarios in the sample. Scenarios frequently describe the fate of individual content creators that are pushed out of their job because of the use of generative AI. Consequently, these individual job losses can also be scaled up and address impacts at an organizational or societal level, and result in loss of revenue for (traditional) news organizations or potential unemployment. While job loss is connected to the individual level, the code unemployment addresses job losses at a macro scale in organizations (e.g., reduction of jobs in the newsroom and/or revenue loss) or societally in terms of the labor market (e.g., mass unemployment). Many scenario writers were worried about a threat for jobs in the media sector and described a future, where fewer jobs in the media sector are available as many are replaced by generative AI.

The labor impacts are highly interrelated with the loss of media quality. An often outlined connection is the economic pressure on journalists and/or journalistic organizations that lead to a frequent, and oftentimes unchecked, use of generative AI. This economic pressure coupled with the loss of human intervention in generating news, resulted in various journalistic quality issues that the scenarios described, like the loss of human touch, the influence of tech corporations, or the lack of credibility. Additionally, the economic impact on individuals (e.g., job loss) is also frequently connected to the well-being theme as people suffer from the increased pressure or from financial shortages.

*4.1.3. Autonomy*

Another dimension that was predominantly found at the individual level, but was infrequently mentioned was connected to the relationship between humans and machines and the resulting impact on human autonomy, or independence to act. Specific instances of this sub-code were present in only a few scenarios, with news consumers mentioning loss of orientation and loss of human autonomy as potential negative impacts, while technology developers highlighted loss of control over AI. Impacts regarding the relation of humans and machines could not be found in the content creator scenarios.

*4.1.4. Legal Rights*

Some scenarios described negative impacts on legal rights. This theme contains impacts related to copyright issues, legal actions (e.g., lawsuits against characters), freedom of expression and the lack of regulation. The sub-codes either address the individual or the societal dimension. On the individual level, for example, content creators face the danger of their material being used for generative AI without their consent, which leads to copyright issues. On the other hand, some scenarios described legal actions because of the misuse of generative AI by some characters. On the societal level, a few scenarios outlined a lack of regulation that leads to the uncontrolled use of generative AI and associated other impacts like the spread of fake news. Altogether, legal rights impacts are only present in a few scenarios, but were mentioned in all stakeholder groups.

*4.1.5. Media Quality*

One of the most mentioned impacts concerned the (loss of) media quality. Here, a plethora of different sub-codes emerged, namely accuracy issues, loss of human touch, sensationalism, credibility/authenticity, lack of diversity/bias, clickbait, journalistic integrity, reframing of narratives, attribution, distinction between journalism and ads, lack of fact-checking, explainability, superficiality, over-personalization, ethics, and accountability. Not all different sub-codes were present in all stakeholder groups (see Table 2), but the concern that media quality is endangered through the use of generative AI was present in nearly every scenario. Especially prevalent are concerns that are related to the inaccuracy of generative AI, which can lead to the dissemination of misinformation (which connects also to impact themes of political impact and/or social cohesion). Another topic that was frequently mentioned was the prioritization of easy and clickable news that was pushed with the use of AI. As such, sensationalism was a common concern amongst all stakeholder groups. Furthermore, biased news was also thematized in many scenarios that also relate to impacts on a societal dimension such as discrimination.

We also found that media quality issues emerge due to different reasons. Some scenario writers describe the emergence of media quality issues as a consequence that emerged through the negligence of actors, i.e. characters failed to anticipate negative impacts of generative AI. The causal agency of the AI system, then, can take various forms, for instance, through a lack of accuracy or sensationalism reinforced by the system's training. Some other scenarios see media quality issues as a result of economic competition (e.g., "Emily's struggle began with news organizations, seeking efficiency and cost reduction increasingly relied on AI- generated news articles." [CC S13]) or as a consequence of intentional agency, specifically malicious use, by characters. Thereby, some characters are aware of potential quality issues, but they decide to take the risk nevertheless due to fear of job loss, economic competition, or personal ambition to work more efficiently. In this regard, some scenario writers also sketch moral dilemmas that characters face as they need to weigh up the pros and cons of using generative AI (e.g., "In fact, its [the generative AI; added by the authors] reliability could be very low, which makes Alejandro face a moral dilemma: should he make use of this technology to write his article or should he do his own research even if it takes more time and effort? "[CC S33]).

*4.1.6. Security*

Security impacts were seldom addressed in the scenarios, though when they were mentioned it was the technology developers who did so. We identified hacking and cybersecurity as sub-codes of the theme. In this theme, scenario writers outline potential lacks of safety that media organizations might have and that could be exploited by malicious actors. For instance, some scenarios describe hacker attacks on news organizations that, consequently, result in other impacts like the spread of misinformation.

*4.1.7. Trustworthiness*

On the level of the media system, the central consequence is a discussion about the trustworthiness of the media environment. Most of the scenarios outline that the trustworthiness of the media is reduced through the use of generative AI. This is expressed through the difficulty to discern between facts and fictional news. Other consequences are that people turn away from the news (media fatigue), consume low quality news, or show a tendency to distrust news altogether. Trustworthiness issues are most prominent in the news consumer scenarios, whereas the content creator scenarios focus more on the media quality aspects. Furthermore trustworthiness issues are also highly connected to political impacts, especially to the spread of misinformation.

Trustworthiness issues occur either on an individual or on a societal level. Some scenarios describe how characters (e.g., news consumers) lose their trust in media content due to the spread of disinformation or elaborate on the struggle of characters to discern between generated and human written news articles. Some scenario writers also outline trustworthiness issues at a societal level and speak of an untrustworthy media environment, which is then also connected to political issues like manipulation. Impacts in the trustworthiness theme can, thus, be a result of intentional behavior by specific actors (e.g., political parties), or a consequence of the widespread use or malfunctioning of generative AI (causal agency).

*4.1.8. Political*

One of the most mentioned impact themes relates to the potential for political impact. The spread of fake news and misinformation was a central topic of the scenarios – over all stakeholder groups. Fake news, in the form of deepfakes, or factually false news, are perceived as a prevalent harm and oftentimes embedded in a specific setting, for example, in the context of elections, or news reporting about crises and wars. The spread of fake news is frequently connected to malicious intentions of specific actors that utilize generative AI for their purposes (e.g., "Pedro is cunning and embodies the bitterness of someone who doesn't look at the means to achieve his ends." [TD S25]). While fake news was mostly connected to political issues, we note that some fake news scenarios also pick up on other topics like personal harassment, which is then connected to well-being. The dimension of fake news is also connected to purposeful manipulation (on a political level). In these scenarios, generative AI, mostly in the form of generating fake news, is frequently used to manipulate citizens' behavior according to the will of a (mostly political) actor. Frequently, scenario-writers tie the political misuse of generative AI to right-wing and/or populist political parties and/or campaigns. Further, gaining an opinion monopoly was found as another negative impact consequence for the accelerated use of generative AI.

Again, political impacts can be found on the individual and societal dimension. On the individual dimension, it describes the susceptibility of news consumers to fake news or manipulation. Some scenarios, for example, point out how news consumers are misled by disinformation campaigns. Interestingly, a few scenarios also elaborate on specific population groups (e.g., old/illiterate people) that are highly susceptible to political misuse. Scenario writers also addressed the societal dimension of scaling up individual impacts and speculate, for instance, that political impacts can have consequences for election outcomes.

*4.1.9. Social Cohesion*

Another central theme in the scenarios concerns social cohesion. The widespread use of generative AI, according to the fears of many scenario-writers, will lead to stronger polarization among the public. This polarization is also caused by the spread of fake news and misinformation on the political level. Polarization itself can also lead to real-world conflicts between societal groups. Related societal consequences outlined in the scenarios are a deepening of social divide, mistrust among societal groups, discrimination of minority groups, as well as stronger dissatisfactions within the population. These concerns are present in all stakeholder groups scenarios and are frequently addressed by scenario writers.

The social cohesion theme is mostly addressed at the societal dimension. Scenario writers, for example, describe high level impacts like fractured societies, polarization, hatred among communities, or international tensions. Interestingly, these impacts are mostly described as part of a result of other impacts, such as political and labor impacts that combine elements that, in the end, enable impacts for social cohesion.

Again, these impacts can be part of a consequence that is based on characters' failure to anticipate negative impacts (negligence), for instance, the introduction of novel generative AI technology that promises to deliver news more efficiently and personalized, leads to further polarization. On the other hand, malicious actors can use generative AI to pursue their goals and actively strive for societal tensions. Again, the interplay between intentions of humans and the affordances of generative AI, then, causes the negative impact.

*4.1.10.* Education

Some scenarios also refer to educational impacts; whereas educational impacts were not frequently mentioned in total, they were articulated by every stakeholder group. Scenario writers speak of a loss of critical engagement as well as a lack of general literacy to deal with the negative impact of generative AI. Again, this dimension is related to several other impacts. For example, a lack of literacy leads to people's inability to discern factual and fake news or to the vulnerability for bad journalistic output. A general loss of literacy within the population can also be an outcome of the use of generative AI, as, so the concerns, the quality of journalistic content could decline on a large-scale and, as such, news consumers do not have the possibility to engage with high quality journalistic content.

## 4.2. Governance strategies

Many respondents, while not explicitly asked to do so, mentioned mitigation strategies for the negative impacts of generative AI on the news environment. These mitigation strategies are often linked to well-intentioned characters. A frequent theme among the scenarios, for example, was the brave (investigative) journalist that fought against the spread of misinformation (e.g., "Pablo is deeply motivated by the pursuit of truth and the belief of journalism is a matter of democracy. They strive daily to serve the public interest and hold those in power [accountable]." [TD S19]). Oftentimes, characters also joined forces to mitigate negative impacts of generative AI, for example, through collaboration. Another theme connected to well-intended mitigation was the invention of some technology solutions to strengthen high quality journalism or to prevent harm. Some characters are also internally motivated by normative values as they want to be on the good side, stand in for their beliefs, or protect their families and friends. In addition, some characters are acting out of democratic or public interest, for example, they want to make news more accessible.

For the analysis in this section, we also included the answers to the open questions following the scenarios (See Section 3.1) to develop the codes for mitigation strategies and transparency obligations.

*4.2.1. Mitigation Strategies*

We identified four main codes for the mitigation strategies outlined in the scenarios: technological approaches, collective action to mitigate the negative impact of the use of generative AI, legal actions, and strategies that aim to restore journalistic quality.

Table 3: Mitigation Strategies

| Main Code | Sub-Codes | Example Quotes<br>NC=News Consumer; CC=Content Creator;<br>TD=Technology Developers; S=Scenario Number |
|---|---|---|
| Technological Approaches | - | "After some months, victims worldwide pressured PhotoMorph's creators. The company proposed a solution: users would provide a photo ID to ensure they only altered their own images." (NC S27)<br>"Together they came up with a plan to use the AI's capabilities to find out its own faults and potential mistakes and correct them within itself. Lola also suggested making the technology more capable of using human intuition and empathy which would make the articles so much more enjoyable." (TD S38) |
| Collective Action | *Protest/Social Movements\*; Public Attention; Public Deliberation; Public Education* | "Emily, determined to counter the damaging effects, embarked on a mission to raise awareness about the dangers of unchecked AI-generated content." (CC S13)<br>"The dangers of relying solely on technology for information are recognized by society, which gives rise to heated discussions about ethics, regulation and the responsibility of tech companies." (NC S35) |
| Legal Actions | *Regulation; Lawsuits; Independent Oversight#; Copyright Protection#; Shifting of Power Distribution#; Restriction of Application Areas#; Restriction of Access#* | "And their efforts paid off, as the government enacted strict regulations on AI algorithms, ensuring transparency and user welfare. Thus, little by little, the toxic social media environment began to dissipate, replaced by healthier and more constructive conversations." (NC S5)<br>"The only comfort that she could find was the law that was created after her case - that none project could be downloaded or screenshoted without the customer service approval who were responsible for checking the content." (NC S29) |
| Restoring Journalistic Quality | *Responsible AI Use/Ethical Guidelines; Re-Focusing on Traditional Journalism; Human Oversight/Fact Checking;* | "She worked with journalists, technology developers and policymakers to establish guidelines to ensure the responsible creation of AI-generated content, creating a news environment where technology complemented human expertise rather than obscuring it." (TD S51)<br>"Collaborating with a group of concerned content creators, she created thought-provoking videos and articles to shed light on the erosion of journalistic |

| | *Accountability; Transparency#; Diversity; Collaborations; Training/Literacy#; Creation of New Jobs/Specialization#* | integrity. Together, they engaged with news consumers, emphasizing the importance of critical thinking and media literacy." (CC S13)<br>"Ethan is one of those - a content creator that could easily make his life easier by using AI technology, but that is choosing to be his own authentic, original self." (CC S27) |
|---|---|---|

Note: *not mentioned in the open answers; #only in open answers

The technological approaches that were suggested aim to mitigate (technological) shortcomings of generative AI encompassed a huge variety of techniques like updates and patches, oversight programs, automated fact-checking tools, fine-tuning of not-accurate models (e.g., self-correcting models), banning of hateful content and prompts, fake-news scanners, rigorous testing, and identity verification. Technological approaches to mitigate negative outcomes were by far most prevalent in the scenarios written by the technology developers. Some scenarios outlined in detail how the proposed tools would help in reducing the negative impacts posed by generative AI.

Collective action strategies were mentioned in all stakeholder groups and comprise the sub-codes of protest/social movement, public attention, public deliberation, and education. The most common theme among the scenarios was some form of social movement or protest that emerged as a reaction to malfunctioning generative AI or negative consequences of the use of generative AI (e.g., spread of fake news). Protests can take the form of raising public attention, and manage to involve citizens to exert pressure on technology companies and/or journalistic organizations. But not all scenarios involved public protest; some just referenced a rise of public attention or public deliberation as a more subtle form of public inclusion. Public deliberation, for example, was outlined as a mitigation strategy that highlighted people's discussion about how generative AI should (not) be used. Gaining public attention, as a related code, relates to the attempts of actors to provide information about the negative impacts of generative AI in the public sphere. However, not all of these attempts, as outlined in scenarios, succeed. Public education was further discussed as a strategy to enable citizens and/or stakeholders to critically assess the changing media environment. In this context, some scenario-writers also ascribe responsibility to news consumers. According to some respondents, the audience also has the obligation to act and think critically and not blindly trust the news.

The legal actions outlined in the scenarios can be differentiated within the sub-codes of regulation and lawsuits. This mitigation strategy was mentioned with similar frequency in all three stakeholder groups. In the scenarios, regulation was mostly a result of the effort of well-intentioned characters made to create the conditions for good quality journalism. Connected with societal action, some scenarios described that regulators stepped in and enforced policies that, for example, foster transparency and accountability standards, user welfare, or stricter oversight of organizations. Other scenarios described lawsuits as a measure to counteract the consequences of some of the impacts outlined. For example, lawsuits can be targeted against specific people or groups that used generative AI with malicious intent.

Another frequently mentioned mitigation strategy concerns directly restoring journalistic quality. In this category, we identified sub-codes for responsible AI use, re-focusing on traditional journalism, fact-checking, accountability measures, investing in diversity, human oversight, and collaborations. Responsible AI use describes ways to ensure a thoughtful use of AI, for example, by following ethical guidelines. Several other related themes occur in this regard, like a need for journalists to fact-check their results continuously, ensuring diversity in the workforce or enforcing human oversight in the production of news. Collaboration between technology developers and people working in journalism is oftentimes proposed in the scenarios to restore journalistic quality. Interestingly, collaboration is foremost mentioned in the technology developers' scenarios, who think of collaboration as a fruitful way to mitigate negative AI impacts. In addition, some scenarios also go one step further and propose a return to traditional journalism without the use of generative AI to avoid the negative impacts.

We further analyzed the answers to the open question that we asked after respondents submitted their scenario (*What could be done to mitigate the risks outlined in the scenario?*). We detected a substantial similarity between the codes that already emerged in the scenarios and the open answers (see superscripts in Table 3 that distinguish where codes were mentioned), but also found that some themes were more accentuated in the open answers. The most prevalent mitigation strategies mentioned in the open questions were regulation, the development of ethical guidelines that are connected to a responsible use of generative AI, the need to strengthen public education, and refraining from the use of generative AI at all. Additionally, some new codes emerged: we found more sub-codes in the areas of legal actions and restoring journalistic quality. Interestingly, also new suggestions for governance interventions were made that are not to our knowledge part of the present regulatory discourse around the European AI Act, such as restrictions on access to the technology for vetted personnel only, not using generative AI for particular forms of journalism (e.g., news coverage about political topics and/or war) but also the creation of "an organization with representatives of all the players involved, from companies and civil society, which would define and create a standard for conduct and behavior for AI technology to be developed, trying the best they can to eliminate biases and discrimination" (TD S43).

For legal actions, we identified various calls for independent oversight. Scenarios suggested, for example, an independent oversight organization and/or NGOs. This regulatory body should see to it that generative AI does not lead to detrimental consequences like the spread of fake news. Additionally, some respondents go one step further and plead for a restriction of the use of generative AI either a) by controlling access, i.e. that only some actors are eligible to use this technology, or b) completely banning the use in specific application areas like the news environment. Also on a structural level, a few respondents propose shifting the power balances in a way that profit-oriented corporations are limited in their ability to push their agenda, i.e. are limited by regulatory boundaries. Lastly, copyright protection was mentioned by some respondents; this was especially prevalent in the content creator group where the question of ownership of training data and/or the output is most vital to their daily work.

On the level of restoring journalistic quality, the sub-codes transparency mechanisms, training, and creation of new (specialized) jobs emerged. Transparency was often mentioned as a strategy to ensure the responsible use of generative AI; this code was also often connected to independent oversight and the need for human oversight. Transparency, for example, could be practically achieved with labeling AI generated content (e.g., with watermarks), or a clear communication of the sources and data used for training generative AI. In addition, training of people working in the media environment and the creation of new (fact-checking) jobs were described by a few participants as suitable measures to enhance journalistic integrity.

*4.2.2. Transparency Obligations*

While the open question invited participants to deliberate relatively freely potential mitigation strategies, we were also interested in respondents' opinion on a *specific* policy proposal, namely transparency obligations as proposed in article 52 of the draft EU AI Act [26]. According to the proposed Article 52 of the draft EU AI Act, consumers have a right to know whether they are interacting with an AI as well as a right to know if content has been artificially generated or manipulated.

We evaluated the open answers regarding the expected effectiveness of the transparency obligations to mitigate the risks outlined in their scenarios. We detected seven categories: *high effectiveness*; *partial/conditional effectiveness*; *change of the scenario, but similar outcome*; *unsureness about the effectiveness*; *small effectiveness, but overall not really important*; *no effectiveness at all*; and *not applicable for the proposed scenario*.

Overall, in each group the major share of respondents welcomed transparency obligations and expected some level of effectiveness. In fact, *high effectiveness* was the most frequent answer in every stakeholder group with 13 (34%) answers in the content creator group, 17 (44%) answers in the technology developer, and 13 (39%) answers in the news consumer group. The high effectiveness of the transparency obligations was mostly linked to the prevention of the spread of fake news as such an obligation would help to clearly label automatically generated content and offer people the opportunity to evaluate content more thoroughly. Strengthening news consumers' awareness was thus frequently mentioned in combination with a high effectiveness. It is also connected to enhancing journalistic quality in a way that it is deemed more trustworthy and helps in distinguishing real and fake content. Some respondents also outline that the scenario they described, would not have taken place from the beginning. For example, ensuring transparency would effectively combat the spread of disinformation as content would be clearly identified as artificially generated and, thus, news consumers would not be misled easily.

Additionally, seven respondents (18%) of the content creator, seven respondents of the technology developer (18%) and nine respondents (27%) of the news consumer group rated transparency obligations as partially effective. The common theme of the answers in this category is that transparency is a step in the right direction, but would not solve all harms caused by generative AI. For example, some respondents believe that transparency labeling is not very effective as news consumers get used to it – some of the answers compare it with warnings on cigarette packs that are judged as useful for only some part of the population, but not for all. Others also indicate that this would not keep content creators away from using generative AI as it is still much cheaper and less time intensive than writing own stories. Furthermore, some respondents, while generally welcoming transparency obligations, question the practical enforcement of it. They doubt that it can be usefully implemented and question the enforceability of such an obligation. Other respondents doubt that news consumers perceive labeling as important as they believe that there is a tendency in the public to trust in the output of algorithms anyway.

However, we also found a sizable portion of respondents, who indicated that transparency obligations would not be effective in counteracting the harms of generative AI. This view is most prevalent in the content creator group with eleven mentions (29%), while only seven technology developers (18%) and five news consumers (15%) expressed this view. The transparency obligation is judged as not effective because of several reasons. First, some respondents do not think that news consumers actually notice labeling of AI generated content either because they skip it or are not paying attention to it. This is also related to the argument of getting used to warning labels that was also mentioned in the partial effective answers. Second, respondents doubt that transparency obligations can be practically enforced. Third, some respondents outline that news consumers would just not care about generative AI created content at all. They portray news consumers as mostly passive and without the ability to critically assess news (anymore). Fourth, some respondents remark that malicious actors simply would not care about transparency obligations and will continue to exploit generative AI for spreading fake news or misinformation. This is also connected to the lack of enforcement that some respondents mention.

Besides the former answers, some respondents answered that they were not sure about the effectiveness of the transparency obligation or report that it would not matter for their scenario, for example, because it is already transparent that generative AI was used. In addition, some respondents believe that the transparency requirement would change the scenario, but nevertheless lead to the same outcome, for example, because malicious actors act the way they do regardless. Finally, a few respondents ascribe the transparency obligation only a small effect, but at the same time doubt that it will have a positive long-term effect.

## 5. DISCUSSION

In the paper, we develop and refine a method to anticipate impacts as well as mitigation strategies for the use of generative AI in the news environment. By inviting different groups of stakeholders to anticipate future impacts of a particular technology, the impact assessment benefits from the insights, expertise and situated experiences of different groups in society. In so doing our more participatory method provides an alternative to predominant methods of impact assessment that are expert driven or grounded in literature reviews of established impacts. In this context, scenario writing is a tool to trigger engagement and reflection as well as sharing participants' own perspectives. Besides identifying new themes of impacts and differences between the kinds of impacts and mitigation strategies between the different stakeholder groups, the method also produces information about the causes of negative impacts in the form of character agencies, i.e. elaborating on stakeholders intentions that may lead to a specific impact. By helping to map the space of action and agency in scenarios the method further sets the stage for future ethics work such as downstream responsibility analysis [28] or the discussion of varying mitigation strategies per stakeholder group. In the following, we will discuss how scenario writing can serve as an impact assessment tool as well as a tool in current governance approaches or policy development. In addition, we will discuss the limitations and propose further research that could be built on our study.

**5.1. Scenario Writing as Impact Assessment tool**

Our study provides rich insights into individual stakeholders' anticipations of the negative impact of generative AI on the news environment. Generative AI already has far-reaching impacts on individuals and society, which will increase even further in the future. Identifying potential negative impacts – and with that informing anticipatory governance – can help in developing strategies on how to prevent this harm as it is easier and more affordable to implement changes earlier in the development and implementation process of emerging technologies. What is more, our method taps into the perceptions and anticipations of the public and, thus, engages a unique perspective that is currently underrepresented in AI impact assessment literature. There, the focus is predominantly on expert- or literature-led approaches [68]. As such, our findings provide valuable insights into a broader societal perspective on generative AI [34]. Especially due to the different roles of the respondents in regard to their interaction with the news environment, either as news consumer, technology developer, or content creator, the findings revealed a variety of perspectives and identification of impacts, but also mitigation strategies, indicating that risks but also mitigation strategies are not one-size-fits-all. As aimed for in the AI impact assessment literature [48,52], the findings also illustrate that use of generative AI is clearly not only and maybe not even most predominantly about individual harms, but also societal harms – whereas the societal harms are far less subject to regulation [66]. Furthermore scenario-writing enables us to tap into the socio-technical interplay [52] and identify impacts that would otherwise be opaque [3]. Thus, an important contribution of our approach is that we expand from a predominantly technical focus on the technology itself to anticipations on how this technology could actually be used in real-world settings by various stakeholder groups.

Compared to other existing AI impact assessment frameworks, we were able to identify some similarities (e.g. references to (mis)information harms, the role of different forms of agency including intentional misuse, security issues, etc.), but also some new aspects that were not identified beforehand. For example, current AI impact assessment of LLMs [71] and text-to-image technology [8] have a strong focus on societal and consumer impacts, but with the inclusion of perspectives of content creators and technology developers, we were also able to identify negative impacts that focus on the economic situation and the corresponding moral trade-offs of people actively working in this sector. In addition, in contrast to the related work on impact assessment, the impacts that were apparent in the scenarios were far more contextually meaningful, getting at quite *specific* ways that generative AI could impact media from personal well-being implications and the need to think about education, to trustworthiness, political implications, and broader concerns of social cohesion. Rather than examine only what impacts technology causes on people, the method allows for a fuller exploration of the sociotechnical interactions where impacts can arise (e.g., labor considerations, social cohesion). We suggest that the method developed can be tuned both based on the instructions and scope of the scenario writing, as well as the specific sample of participants, in order to get far more fine-grained and contextually meaningful impacts than is apparent from methods relying only on experts.

While existing impact assessment methods tend to focus on mapping the potential risks, using our method we were able to illuminate the rationale and motivations of actors using and responding to generative AI – again highlighting the socio-technical perspective that this method brings to the table [52]. This offers the potential to also inform an eventual analysis of agency or responsibility, whether ethical or for informing formal regulatory approaches. As shown in the findings, characters in the media environment exhibited different types of agency. In the triadic agency sense, generative AI is only the tool (causal agency) that contributes to a negative impact. However, impacts always trace back to the intentions of characters. To discuss ethical responsibility, focusing on the different forms of intentions, then, makes a difference. For example, we found that characters in the scenarios acted out of malicious behavior (e.g., purposefully spreading fake news), but other impacts emerged mainly due to negligence (e.g., characters failed to anticipate technological malfunctioning), or describe the complex interplay between characters and generative AI (e.g., economic pressure on organizations/journalists lead to the use of generative AI that causes negative impacts). Some scenarios made also apparent the moral struggles that the use of generative AI can bring with it. The scenario writing method therefore offers specific examples that can help fuel an (ethical) discussion about agency and responsibility.

The scenarios offer vivid examples of unique perspectives that are currently missing in high-level description and categorization schemes of AI impact assessments. These unique perceptions are also based on the concrete role that individuals take in the news environment. This can be traced back to the different roles these groups take in regard to generative AI.

For news consumers, impacts regarding well-being are more prevalent as they imagine their characters in a user-setting, in which they describe how individuals interact with the emerging technology. Likewise, news consumers are concerned about the trustworthiness of the media content they consume. Technology developers bring their unique perspective when it comes to the articulation of safety issues. This is a rather technical perspective, which relates to their profession on how to build generative AI systems that are trustworthy and safe to use. This impact category is therefore also role-specific. Content creators also frequently highlight impacts that are related to their profession. We showed that content creators were overall concerned about generative AI's impact on the media quality, highlighting diverse specific issues that could be endangered by the use of generative AI. As professionals, their scenarios describe unique and diverse possible impacts that could not be found (in such richness) within the scenarios of the other stakeholder groups. As this diverging presence of codes in each stakeholder group shows, news consumers, technology developers, and content creators raise awareness on different impacts, thus highlighting the value of sampling for cognitive diversity and a range of expertise and thus supporting the participatory foresight approach utilized in this study [12,54]. At the same time, we also acknowledge that some impact categories were equally mentioned in all stakeholder groups. These impact categories can be seen as issues of common concern that respondents are aware of regardless of their role in relation to the use of generative AI in the news environment. Those impact themes are labor, politics, legal rights, education, and social cohesion. This can be explained by the specific topics that these categories entail. As, for instance, disinformation and the potential threat for jobs are highly discussed in the public debate, it is not surprising that all stakeholder groups thematize these impacts. In addition, issues such as legal rights, education, and social cohesion are themes that are not specific to stakeholder's roles and are, thus, mentioned equally across the groups.

At the same time as we see benefits to this method we also suggest there is still value in expert-driven impact assessment tools as some impacts identified by experts were not found in the scenarios in our sample. For example, the environmental costs of generative AI [9,18,35,67,71] are not mentioned in any of the scenarios, which corresponds to studies that highlight the unawareness of citizens and the public debate with the environmental costs of AI [1,42,44]. This also exposes a limitation of the method insofar as our task description specifically oriented respondents' attention towards impacts on the media ecosystem, and in this case no respondent saw the connection to environmental concerns. Our approach can thus be seen as complementary to existing expert-led approaches and also subject to how the task is framed and presented to respondents.

## 5.2. Scenario-Writing for Impact Assessments and Policy Development

Our study also demonstrated that scenario-writing can help not only in identifying individual and societal impacts but also mitigation strategies – a crucial component of anticipatory governance research [34,63] and AI impact assessment [3,52]. The fact that our respondents discussed mitigation strategies even without being actively asked to do so demonstrates how the scenario-writing exercise resulted in active engagement and stimulated critical thinking in the respondents. This also suggests the potential of using scenario-writing not only for negative impact identification, but to help develop mitigation strategies that are grounded in the experience of different affected stakeholder groups and that can again inform and inspire policy options. Any policy intervention is only as good as its enforceability, and information on what mitigation strategies stakeholders themselves consider effective can inform policy makers on what mitigation strategies are more likely to be supported 'on the ground'. This could serve as a valuable addition to the usually expert driven and top-down methods to develop mitigation strategies. We identified technological approaches, collective action, legal action, and restoring journalistic quality as overlying mitigation strategies. Again, we could find some differences between the respondents' role in regard to generative AI. Technology developers brought their professional experience in and highlighted technological fixes for malfunctioning generative AI systems as well as a strengthening of collaboration between journalists and technological experts. On the other hand, content creators emphasized copyright protection as an important mitigation strategy. As these findings show, personal and professional experiences informed also the proposal of mitigation strategies. Thus, sampling for cognitive diversity provided not only useful insights for the identification of impacts, but also for how these impacts could be mitigated.

The findings in our study also revealed novel ideas to mitigate harms that are currently not discussed prominently in existing policy debates around the AI Act. Examples for such policy options include the restriction of access to generative AI only for vetted personnel and the creation of an oversight organization for generative AI use in the news environment consisting of representatives of different stakeholder groups. The responses also made clear that there is not one single mitigation strategy, but that in order to mitigate the potential negative impacts from AI, a combination of different strategies is needed and reflects the societal complexity in which generative AI functions. Our approach and findings thus contribute to the identification of "reasonably foreseeable risks", as called for in the draft EU AI-Act. Policy-makers have already deployed scenario-writing as a method to anticipate the impacts of new technologies on society and/or specific domains like journalism [25,39]. We add an academic perspective to the current approaches and utilize a diverse sampling to provide even more information for political decision-makers. The scenarios and the mitigation strategies can serve as a starting point for impact assessments, and to engage actively with policy-makers about possible mitigation measures. The added contribution of this particular paper is testing scenario writing as a form of bottom-up impact mapping, and as the basis for future work on a critical analysis of existing regulatory approaches. The work is also useful in identifying the breadth of possible mitigation strategies and how these may differ between stakeholder groups. Future work could follow up with (selected) scenarios and/or mitigation strategies and discuss to what extent current governance approaches already address the concerns and proposed mitigation strategies, or where doing so could be a viable route for future policy development. For instance, our findings hint at policy areas, which are important for many respondents. Here, especially the strong emphasis on economic consequences for affected stakeholders can be highlighted; an issue, while recently addressed by some political decision makers and policy white papers [e.g., 7], that is still less prominent in current policy debates and in scholarly research on generative AI ethics than other issues (e.g., fairness, safety, or toxicity) [36]. Consequently, this seems to be a crucial dimension of ethical impacts that needs to be addressed in future policy discussion.

It is interesting that individuals in the scenarios focus on collective social action as a mitigation strategy, whereas regulatory approaches, such as the AI Act focus much more on individual rights and protected interests, and far less on how to enable collective action or involvement of civil society. Insights like these beg the question how AI governance approaches could support collective action as a counterweight to the power and potential of AI in society. Utilizing scenario-writing to tap into lived realities of affected individuals as a means of revisiting some existing policy debates might thus be a useful practical addition to inform policy debates.

The transparency questions yielded further insights into how a specific policy action could help (or not) in mitigating impacts caused by AI. While there is a general tendency to approve and welcome the transparency obligation, content creators are a bit more skeptical about its effectiveness than respondents of the other groups. This skepticism is attributed to a lack of enforcement mechanism, or a general disbelief in the capabilities of the public to actually pay attention to transparency measures. All in all, the open answers to this question provide additional insights that can help decision makers deliberate about implementing and operationalizing transparency mechanisms. Furthermore, the variety of governance approaches that were hinted at as respondents talked about mitigation strategies could be subjected to the same kind of targeted evaluation as we did here for transparency as part of future work.

### 5.3. Limitations & Outlook

There are some limitations that have to be acknowledged. First, our sampling strategy was aimed to ensure diversity in regard to respondents' country of residence in the EU and sex distribution. However, our sample ultimately consisted predominantly of people who identify themselves as White and well-educated. As frequently pointed out in the scholarly debate about access and participation, voices from marginalized communities are direly needed as they bring in unique perspectives and raise awareness on aspects that are often overseen or neglected by the industry, developers or some scholars [18,52,59]. However, public opinion research also reports that it is difficult to reach these groups as they are particularly uninvolved in the public discourse on AI [4,43]. Future research should develop approaches that particularly aim to include the voices of those communities that are usually not well presented in the public debate as well as the scholarly debate about AI impacts. This could be achieved through targeted sampling in surveys with additional attention given to sampling on dimensions of factors such as gender identity and race, or workshops developed in collaboration with NGOs and interest groups.

Additionally, further research should develop approaches to validate and evaluate the findings of the scenarios. As of now, scenarios are tied to the imagination of the respondents and are not necessarily fully plausible from a technical, legal, or societal viewpoint. For example, a total ban of generative AI tools may conflict with citizens' fundamental right to freedom of expression or economic freedoms. Banning generative AI altogether is, thus, not a viable suggestion. Consequently, validation and synthesis is a crucial next step to make scenarios useful for the policy debate about impact mitigation. Further, besides validation, the scenarios can also be evaluated in terms of different variables of interest. For instance, an insightful further dimension for impact assessment would be to rate the severity of the impacts or the likelihood of occurrence. This could be achieved through (expert) workshops, or Delphi studies that synthesize multiple expertise profiles to assess viability. Another option would be to conduct a quantitative survey among news consumers (or stakeholder groups) and let them evaluate other people's scenarios or refined scenarios that are carefully constructed based on our findings, respectively. A quantitative validation could aim for different aspects such as plausibility of the scenarios, severity of impacts, likelihood of an impact materializing, but also (individual/societal) desirability of scenarios or easiness to understand. Further, also proposed policy options that emerged as a result of this study could be evaluated for their feasibility and effectiveness. This approach was recently used by Dobber et al. for the use case of veracity labeling in political advertising [23].

Further research can build on our groundwork and dig deeper into the notion of cognitive diversity. Relying on studies of AI narratives and imaginaries lead to the assumption that socio-demographic and AI-related factors have an influence on the emerging scenarios [16,18,40,59]. Tapping deeper into the underlying factors that influence the creation of the scenarios can further provide a better understanding of possible risk dynamics but also result in more diversity in the results – and better position the voices of particular stakeholder or minority groups. Anecdotal findings from our scenarios indeed point out that these factors matter in terms of scenario-writing: For example, we found that respondents from Poland thematized Russia's war of aggression on Ukraine in violation of international law and, in light of this, connected AI impacts to this topic, e.g. the rise and spread of disinformation in a political setting. However, our sample size is not suitable for a systematic and thorough analysis between respondents with different socio-demographic information (e.g., country of residence). Exploring those differences is a promising research avenue for future scholarly work.

In a connected world, the anticipation of generative AI's impact in the news ecosystem is a global challenge. Especially in light of the potential negative impacts of generative AI on elections (like the upcoming US and EU elections in 2024), we need knowledge on how potential impacts can unfold and – even more importantly – how they could be addressed. Thus, there is a need to update and expand our approach to the global scale. Including residents of other countries beyond the EU might result in the identification of different impact classifications since political, cultural, and socio-economic factors influence the perspectives and imaginations of generative AI's impact in these countries as well as perceptions regarding AI technology [41]. Anticipating impacts is also relevant in light of the emerging regulatory frameworks like the EU AI Act, Biden's Executive Order on AI, but also for countries that are in the process of developing legal frameworks on AI. Identifying potential detrimental impacts of generative AI such as those presented in this study could inform policy-making processes and shine light on issues that need to be addressed by legislators. These strategies, however, should also be informed by studies with residents of the respective countries. Our study offers an approach that can be applied by researchers as well as political decision makers tasked with developing governance strategies.

## 6. CONCLUSION

In this work we systematically anticipated and mapped the impacts of generative AI as well as corresponding mitigation strategies and a concrete policy proposal currently under discussion, namely transparency obligations as outlined in Article 52 of the draft EU AI Act. In applying scenario-writing we delve into the cognitively diverse future imaginations of news consumers, technology developers, and content creators. Our findings show that scenario writing via diverse sampling on a survey platform is a promising approach for anticipating the impact of generative AI and related mitigation strategies. Further, different stakeholder groups raise awareness of a variety of potential impacts based on their own unique perspectives and expertise. In detail, we identified ten impact themes with fifty specific impacts, whereby the negative impacts of generative AI on media quality as well as economic impacts dominated. In regard to mitigation strategies, we identified four main categories with twenty specific strategies, including some that were novel to existing governance strategies. In addition, transparency obligations are seen as a viable measure to address some of the potential harms of generative AI.

## FUNDING

The funding for this research was provided by UL Research Institutes through the Center for Advancing Safety of Machine Intelligence.

## CONFLICT OF INTEREST

The authors declare no competing interests.

## REFERENCES

[1] Sarah Akyürek, Kimon Kieslich, Pero Dosenovic, Frank Marcinkowski, and Esther Laukötter. 2022. Environmental Sustainability of Artificial Intelligence. How does the public perceive the environmental footprint of artificial intelligence? (2022). https://doi.org/10.13140/RG.2.2.33348.09600

[2] Muhammad Amer, Tugrul U. Daim, and Antonie Jetter. 2013. A review of scenario planning. *Futures : the journal of policy, planning and futures studies* 46, (February 2013), 23–40. https://doi.org/10.1016/j.futures.2012.10.003

[3] Adam Amos-Binks, Dustin Dannenhauer, and Leilani H. Gilpin. 2023. The anticipatory paradigm. *AI Magazine* 44, 2 (June 2023), 133–143. https://doi.org/10.1002/aaai.12098

[4] Luye Bao, Nicole M. Krause, Mikhaila N. Calice, Dietram A. Scheufele, Christopher D. Wirz, Dominique Brossard, Todd P. Newman, and Michael A. Xenos. 2022. Whose AI? How different publics think about AI and its social impacts. *Computers in human behavior* 130, (May 2022), 107182. https://doi.org/10.1016/j.chb.2022.107182

[5] Julia Barnett and Nicholas Diakopoulos. 2022. Crowdsourcing Impacts: Exploring the Utility of

Crowds for Anticipating Societal Impacts of Algorithmic Decision Making. In *Proceedings of the 2022 AAAI/ACM Conference on AI, Ethics, and Society*, July 26, 2022, Oxford United Kingdom. ACM, Oxford United Kingdom, 56–67. . https://doi.org/10.1145/3514094.3534145

[6]  Charlie Beckett and Mira Yaseen. 2023. *Generating Change. A global survey of what news organisations are doing with artificial intelligence.* Retrieved from https://static1.squarespace.com/static/64d60527c01ae7106f2646e9/t/6509b9a39a5ca70df9148eac/1695136164679/Generating+Change+_+The+Journalism+AI+report+_+English.pdf

[7]  Janine Berg, Mark Graham, Marek Havrda, Matthias Peissner, Saiph Savage, Basheerhamad Shadrach, Fernando Schapachnik, Alexandre Shee, Lucía Velasco, and Kyoko Yoshinaga. 2023. *Policy Brief: Generative AI, Jobs, and Policy Response*. The Global Partnership on Artificial Intelligence. Retrieved from https://media.licdn.com/dms/document/media/D4E1FAQGPh3WfCMxQWw/feedshare-document-pdf-analyzed/0/1696184236735?e=1697673600&v=beta&t=Wl-xE3w2RWez20YBgRA4je5vdHd5oY5oHRtS-Nyv6ZY

[8]  Charlotte Bird, Eddie L. Ungless, and Atoosa Kasirzadeh. 2023. Typology of Risks of Generative Text-to-Image Models. Retrieved August 10, 2023 from http://arxiv.org/abs/2307.05543

[9]  Rishi Bommasani, Drew A. Hudson, Ehsan Adeli, Russ Altman, Simran Arora, Sydney von Arx, Michael S. Bernstein, Jeannette Bohg, Antoine Bosselut, Emma Brunskill, Erik Brynjolfsson, Shyamal Buch, Dallas Card, Rodrigo Castellon, Niladri Chatterji, Annie Chen, Kathleen Creel, Jared Quincy Davis, Dora Demszky, Chris Donahue, Moussa Doumbouya, Esin Durmus, Stefano Ermon, John Etchemendy, Kawin Ethayarajh, Li Fei-Fei, Chelsea Finn, Trevor Gale, Lauren Gillespie, Karan Goel, Noah Goodman, Shelby Grossman, Neel Guha, Tatsunori Hashimoto, Peter Henderson, John Hewitt, Daniel E. Ho, Jenny Hong, Kyle Hsu, Jing Huang, Thomas Icard, Saahil Jain, Dan Jurafsky, Pratyusha Kalluri, Siddharth Karamcheti, Geoff Keeling, Fereshte Khani, Omar Khattab, Pang Wei Koh, Mark Krass, Ranjay Krishna, Rohith Kuditipudi, Ananya Kumar, Faisal Ladhak, Mina Lee, Tony Lee, Jure Leskovec, Isabelle Levent, Xiang Lisa Li, Xuechen Li, Tengyu Ma, Ali Malik, Christopher D. Manning, Suvir Mirchandani, Eric Mitchell, Zanele Munyikwa, Suraj Nair, Avanika Narayan, Deepak Narayanan, Ben Newman, Allen Nie, Juan Carlos Niebles, Hamed Nilforoshan, Julian Nyarko, Giray Ogut, Laurel Orr, Isabel Papadimitriou, Joon Sung Park, Chris Piech, Eva Portelance, Christopher Potts, Aditi Raghunathan, Rob Reich, Hongyu Ren, Frieda Rong, Yusuf Roohani, Camilo Ruiz, Jack Ryan, Christopher Ré, Dorsa Sadigh, Shiori Sagawa, Keshav Santhanam, Andy Shih, Krishnan Srinivasan, Alex Tamkin, Rohan Taori, Armin W. Thomas, Florian Tramèr, Rose E. Wang, William Wang, Bohan Wu, Jiajun Wu, Yuhuai Wu, Sang Michael Xie, Michihiro Yasunaga, Jiaxuan You, Matei Zaharia, Michael Zhang, Tianyi Zhang, Xikun Zhang, Yuhui Zhang, Lucia Zheng, Kaitlyn Zhou, and Percy Liang. 2021. On the Opportunities and Risks of Foundation Models. (2021). https://doi.org/10.48550/ARXIV.2108.07258

[10] Andrea Bonaccorsi, Riccardo Apreda, and Gualtiero Fantoni. 2020. Expert biases in technology foresight. Why they are a problem and how to mitigate them. *Technological forecasting & social change* 151, (February 2020). https://doi.org/10.1016/j.techfore.2019.119855

[11] Lena Börjeson, Mattias Höjer, Karl-Henrik Dreborg, Tomas Ekvall, and Göran Finnveden. 2006. Scenario types and techniques: Towards a user's guide. *Futures* 38, 7 (2006), 723–739. https://doi.org/10.1016/j.futures.2005.12.002

[12] Philip Brey. 2017. Ethics of emerging technology. *The ethics of technology: Methods and approaches* (2017), 175–191.

[13] Philip A. E. Brey. 2012. Anticipatory Ethics for Emerging Technologies. *Nanoethics* 6, 1 (April 2012), 1–13. https://doi.org/10.1007/s11569-012-0141-7

[14] Zana Buçinca, Chau Minh Pham, Maurice Jakesch, Marco Tulio Ribeiro, Alexandra Olteanu, and Saleema Amershi. 2023. AHA!: Facilitating AI Impact Assessment by Generating Examples of

[15] Michael Burnam-Fink. Creating narrative scenarios: Science fiction prototyping at Emerge. https://doi.org/10.1016/j.futures.2014.12.005

[16] Stephen Cave, Claire Craig, Kanta Dihal, Sarah Dillon, Jessica Montgomery, Beth Singler, and Lindsay Taylor. 2018. *Portrayals and perceptions of AI and why they matter*. Apollo - University of Cambridge Repository. https://doi.org/10.17863/cam.34502

[17] Alan Chan, Rebecca Salganik, Alva Markelius, Chris Pang, Nitarshan Rajkumar, Dmitrii Krasheninnikov, Lauro Langosco, Zhonghao He, Yawen Duan, Micah Carroll, Michelle Lin, Alex Mayhew, Katherine Collins, Maryam Molamohammadi, John Burden, Wanru Zhao, Shalaleh Rismani, Konstantinos Voudouris, Umang Bhatt, Adrian Weller, David Krueger, and Tegan Maharaj. 2023. Harms from Increasingly Agentic Algorithmic Systems. (2023). https://doi.org/10.48550/ARXIV.2302.10329

[18] Kate Crawford. 2021. *The atlas of AI: Power, politics, and the planetary costs of artificial intelligence*. Yale University Press.

[19] Erik De Vries, Martijn Schoonvelde, and Gijs Schumacher. 2018. No Longer Lost in Translation: Evidence that Google Translate Works for Comparative Bag-of-Words Text Applications. *Polit. Anal.* 26, 4 (October 2018), 417–430. https://doi.org/10.1017/pan.2018.26

[20] Nicholas Diakopoulos. 2020. Computational News Discovery: Towards Design Considerations for Editorial Orientation Algorithms in Journalism. *Digital Journalism* 8, 7 (August 2020), 945–967. https://doi.org/10.1080/21670811.2020.1736946

[21] Nicholas Diakopoulos and Deborah Johnson. 2021. Anticipating and addressing the ethical implications of deepfakes in the context of elections. *New Media & Society* 23, 7 (2021), 2072–2098. https://doi.org/10.1177/1461444820925811

[22] Nick Diakopoulos. 2023. The State of AI in Media: From Hype to Reality. *Medium*. Retrieved August 21, 2023 from https://generative-ai-newsroom.com/the-state-of-ai-in-media-from-hype-to-reality-37b250541752

[23] Tom Dobber, Sanne Kruikemeier, Fabio Votta, Natali Helberger, and Ellen P. Goodman. 2023. The effect of traffic light veracity labels on perceptions of political advertising source and message credibility on social media. *Journal of Information Technology & Politics* (June 2023), 1–16. https://doi.org/10.1080/19331681.2023.2224316

[24] European Commission. 2021. *Proposal for a regulation of the European Parliament and of the Council of laying down harmonised rules on artificial intelligence (Artificial Intelligence Act) and amending certain union legislative acts*.

[25] European Commission. Joint Research Centre. 2023. *Reference foresight scenarios on the global standing of the EU in 2040*. Publications Office, LU. Retrieved October 18, 2023 from https://data.europa.eu/doi/10.2760/490501

[26] European Parliament. 2023. *Texts adopted - Artificial Intelligence Act - Wednesday, 14 June 2023*. Retrieved August 9, 2023 from https://www.europarl.europa.eu/doceo/document/TA-9-2023-0236_EN.html

[27] eurostat. 2022. *New indicator on annual average salaries in the EU*. Retrieved from https://ec.europa.eu/eurostat/web/products-eurostat-news/w/ddn-20221219-3

[28] Jessica Nihlén Fahlquist. 2017. Responsibility analysis. *The Ethics of Technology. Methods and Approaches* (2017), 129–143.

[29] Leon Fuerth. 2011. Operationalizing Anticipatory Governance. *PRISM* 2, 4 (2011), 31–46.

[30] Tarleton Gillespie. 2020. Content moderation, AI, and the question of scale. *Big Data & Society* 7, 2 (July 2020), 205395172094323. https://doi.org/10.1177/2053951720943234

[31] Barney Glaser and Anselm Strauss. 2017. *Discovery of grounded theory: Strategies for qualitative research*. Routledge.

[32] Michel Godet. 2000. How to be rigorous with scenario planning. *Foresight* 2, 1 (February 2000), 5–9. https://doi.org/10.1108/14636680010802438

[33] Robert Gorwa, Reuben Binns, and Christian Katzenbach. 2020. Algorithmic content moderation: Technical and political challenges in the automation of platform governance. *Big Data & Society* 7, 1 (January 2020), 205395171989794. https://doi.org/10.1177/2053951719897945

[34] Guston. 2013. Understanding 'anticipatory governance.' *Social Studies of Science* 44, 2 (2013), 218–242. https://doi.org/10.1177/0306312713508669

[35] Philipp Hacker. 2023. Sustainable AI Regulation. *SSRN Journal* (2023). https://doi.org/10.2139/ssrn.4467684

[36] Thilo Hagendorff. 2024. Mapping the Ethics of Generative AI: A Comprehensive Scoping Review. Retrieved February 20, 2024 from http://arxiv.org/abs/2402.08323

[37] Mia Hoffmann and Heather Frase. 2023. *Adding Structure to AI Harm*. Center for Security and Emerging Technology. Retrieved July 31, 2023 from https://cset.georgetown.edu/publication/adding-structure-to-ai-harm/

[38] Deborah G. Johnson and Mario Verdicchio. 2019. AI, agency and responsibility: the VW fraud case and beyond. *AI & Soc* 34, 3 (September 2019), 639–647. https://doi.org/10.1007/s00146-017-0781-9

[39] Ila Kasem, Mark van Waes, and Kim Wannet. 2015. *What's New(s)? Scenarios for the future of journalism*. Stimuleringsfonds voor de Journalistiek. Retrieved from https://www.journalism2025.com/bundles/svdjui/documents/Scenarios-for-the-future-of-journalism.pdf

[40] Christian Katzenbach. 2021. "AI will fix this" – The Technical, Discursive, and Political Turn to AI in Governing Communication. *Big data & society* 8, 2 (July 2021), 205395172110461. https://doi.org/10.1177/20539517211046182

[41] Patrick Gage Kelley, Yongwei Yang, Courtney Heldreth, Christopher Moessner, Aaron Sedley, Andreas Kramm, David T. Newman, and Allison Woodruff. 2021. Exciting, Useful, Worrying, Futuristic: Public Perception of Artificial Intelligence in 8 Countries. In *Proceedings of the 2021 AAAI/ACM Conference on AI, Ethics, and Society*, July 21, 2021, Virtual Event USA. ACM, Virtual Event USA, 627–637. . https://doi.org/10.1145/3461702.3462605

[42] Kimon Kieslich, Pero Došenović, and Frank Marcinkowski. 2022. *Everything, but hardly any science fiction*. Meinungsmonitor Künstliche Intelligenz. Retrieved from https://www.researchgate.net/profile/Kimon-Kieslich/publication/365033703_Everything_but_hardly_any_science_fiction/links/63638442431b1f5300685b2d/Everything-but-hardly-any-science-fiction.pdf

[43] Kimon Kieslich, Marco Lünich, and Pero Došenović. 2023. Ever Heard of Ethical AI? Investigating the Salience of Ethical AI Issues among the German Population. *International Journal of Human–Computer Interaction* (February 2023), 1–14. https://doi.org/10.1080/10447318.2023.2178612

[44] Pascal D König, Stefan Wurster, and Markus B Siewert. 2022. Consumers are willing to pay a price for explainable, but not for green AI. Evidence from a choice-based conjoint analysis. *Big Data & Society* 9, 1 (January 2022), 205395172110696. https://doi.org/10.1177/20539517211069632

[45] Fabienne Lind, Jakob-Moritz Eberl, Olga Eisele, Tobias Heidenreich, Sebastian Galyga, and Hajo G. Boomgaarden. 2022. Building the Bridge: Topic Modeling for Comparative Research. *Communication Methods and Measures* 16, 2 (April 2022), 96–114. https://doi.org/10.1080/19312458.2021.1965973

[46] John Lofland, David Snow, Leon Anderson, and Lyn H. Lofland. 2022. *Analyzing social settings: A guide to qualitative observation and analysis*. Waveland Press.

[47] Anna-Katharina Meßmer and Martin Degeling. 2023. *Auditing Recommender Systems. Putting the DSA into practice wit a risk-scenario-based approach*. Stiftung Neue Verantwortung. Retrieved from https://www.stiftung-nv.de/sites/default/files/auditing.recommender.systems.pdf

[48] Jacob Metcalf, Emanuel Moss, Elizabeth Anne Watkins, Ranjit Singh, and Madeleine Clare Elish.

[] 2021. Algorithmic impact assessments and accountability: the co-construction of impacts. *Proceedings of the 2021 ACM Conference on Fairness, Accountability, and Transparency* (2021). https://doi.org/10.1145/3442188.3445935

[49] Yisroel Mirsky, Ambra Demontis, Jaidip Kotak, Ram Shankar, Deng Gelei, Liu Yang, Xiangyu Zhang, Maura Pintor, Wenke Lee, Yuval Elovici, and Battista Biggio. 2023. The Threat of Offensive AI to Organizations. *Computers & Security* 124, (January 2023), 103006. https://doi.org/10.1016/j.cose.2022.103006

[50] Brent Daniel Mittelstadt, Bernd Carsten Stahl, and N. Ben Fairweather. 2015. How to Shape a Better Future? Epistemic Difficulties for Ethical Assessment and Anticipatory Governance of Emerging Technologies. *Ethic Theory Moral Prac* 18, 5 (November 2015), 1027–1047. https://doi.org/10.1007/s10677-015-9582-8

[51] Shakir Mohamed, Marie-Therese Png, and William Isaac. 2020. Decolonial AI: Decolonial Theory as Sociotechnical Foresight in Artificial Intelligence. *Philos. Technol.* 33, 4 (December 2020), 659–684. https://doi.org/10.1007/s13347-020-00405-8

[52] Emanuel Moss, Elizabeth Watkins, Ranjit Singh, Madeleine Clare Elish, and Jacob Metcalf. 2021. Assembling Accountability: Algorithmic Impact Assessment for the Public Interest. *SSRN Journal* (2021). https://doi.org/10.2139/ssrn.3877437

[53] Priyanka Nanayakkara, Nicholas Diakopoulos, and Jessica Hullman. 2020. Anticipatory ethics and the role of uncertainty. *arXiv preprint arXiv:2011.13170* (2020).

[54] Blagovesta Nikolova. 2014. The rise and promise of participatory foresight. *European Journal of Futures Research* 2, 1 (2014). https://doi.org/10.1007/s40309-013-0033-2

[55] Sanchita Nishal and Nicholas Diakopoulos. 2023. Envisioning the Applications and Implications of Generative AI for News Media. (2023).

[56] Ray Quay. 2010. Anticipatory Governance: A Tool for Climate Change Adaptation. *Journal of the American Planning Association* 76, 4 (September 2010), 496–511. https://doi.org/10.1080/01944363.2010.508428

[57] Rafael Ramírez and Cynthia Selin. 2014. Plausibility and probability in scenario planning. *Foresight (Cambridge)* 16, 1 (January 2014), 54–74. https://doi.org/10.1108/FS-08-2012-0061

[58] Tore G. C. Rich. 2023. Document Summaries in Danish with OpenAI. *Medium*. Retrieved June 8, 2023 from https://generative-ai-newsroom.com/summaries-in-danish-with-openai-cbb814a119f2

[59] Laura Sartori and Andreas Theodorou. 2022. A sociotechnical perspective for the future of AI: narratives, inequalities, and human control. *Ethics and information technology* 24, 1 (March 2022). https://doi.org/10.1007/s10676-022-09624-3

[60] Adam Satariano and Paul Mozur. 2023. The People Onscreen Are Fake. The Disinformation Is Real. *The New York Times*. Retrieved August 21, 2023 from https://www.nytimes.com/2023/02/07/technology/artificial-intelligence-training-deepfake.html

[61] Paul J. H. Schoemaker. 1991. When and how to use scenario planning: A heuristic approach with illustration. *J. Forecast.* 10, 6 (November 1991), 549–564. https://doi.org/10.1002/for.3980100602

[62] Elizabeth Seger, Aviv Ovadya, Ben Garfinkel, Divya Siddarth, and Allan Dafoe. 2023. Democratising AI: Multiple Meanings, Goals, and Methods. Retrieved August 10, 2023 from http://arxiv.org/abs/2303.12642

[63] Andrew D. Selbst. 2021. AN INSTITUTIONAL VIEW OF ALGORITHMIC IMPACT. *Harvard Journal of Law & Technology* 35, 1 (2021).

[64] Cynthia Selin. 2006. Trust and the illusive force of scenarios. *Futures : the journal of policy, planning and futures studies* 38, 1 (February 2006), 1–14. https://doi.org/10.1016/j.futures.2005.04.001

[65] Renee Shelby, Shalaleh Rismani, Kathryn Henne, AJung Moon, Negar Rostamzadeh, Paul Nicholas, N'Mah Yilla, Jess Gallegos, Andrew Smart, Emilio Garcia, and Gurleen Virk. 2023. Sociotechnical Harms of Algorithmic Systems: Scoping a Taxonomy for Harm Reduction. Retrieved August 1, 2023

from http://arxiv.org/abs/2210.05791

[66] Nathalie A. Smuha. 2021. Beyond the individual: governing AI's societal harm. *Internet Policy Review* 10, 3 (September 2021). https://doi.org/10.14763/2021.3.1574

[67] Irene Solaiman, Zeerak Talat, William Agnew, Lama Ahmad, Dylan Baker, Su Lin Blodgett, Hal Daumé III, Jesse Dodge, Ellie Evans, Sara Hooker, Yacine Jernite, Alexandra Sasha Luccioni, Alberto Lusoli, Margaret Mitchell, Jessica Newman, Marie-Therese Png, Andrew Strait, and Apostol Vassilev. 2023. Evaluating the Social Impact of Generative AI Systems in Systems and Society. Retrieved June 14, 2023 from http://arxiv.org/abs/2306.05949

[68] Bernd Carsten Stahl, Josephina Antoniou, Nitika Bhalla, Laurence Brooks, Philip Jansen, Blerta Lindqvist, Alexey Kirichenko, Samuel Marchal, Rowena Rodrigues, Nicole Santiago, Zuzanna Warso, and David Wright. 2023. A systematic review of artificial intelligence impact assessments. *Artif Intell Rev* (March 2023). https://doi.org/10.1007/s10462-023-10420-8

[69] Edward Tian and Alexander Cui. 2023. GPTZero: Towards detection of AI-generated text using zero-shot and supervised methods. Retrieved from https://gptzero.me

[70] Veniamin Veselovsky, Manoel Horta Ribeiro, and Robert West. 2023. Artificial Artificial Artificial Intelligence: Crowd Workers Widely Use Large Language Models for Text Production Tasks. https://doi.org/10.48550/arXiv.2306.07899

[71] Laura Weidinger, John Mellor, Maribeth Rauh, Conor Griffin, Jonathan Uesato, Po-Sen Huang, Myra Cheng, Mia Glaese, Borja Balle, Atoosa Kasirzadeh, Zac Kenton, Sasha Brown, Will Hawkins, Tom Stepleton, Courtney Biles, Abeba Birhane, Julia Haas, Laura Rimell, Lisa Anne Hendricks, William Isaac, Sean Legassick, Geoffrey Irving, and Iason Gabriel. 2021. Ethical and social risks of harm from Language Models. Retrieved from http://arxiv.org/pdf/2112.04359v1 http://arxiv.org/abs/2112.04359v1 https://arxiv.org/pdf/2112.04359v1.pdf

[72] Silke Zimmer-Merkle and Torsten Fleischer. 2017. Eclectic, random, intuitive? Technology assessment, RRI, and their use of history. *Journal of Responsible Innovation* 4, 2 (May 2017), 217–233. https://doi.org/10.1080/23299460.2017.1338105

# APPENDIX

## A1: Introduction to Generative AI

Generative AI is a technology that can create new content (e.g. text, images, audio, video) based on the content it was trained on.

**Capabilities**
- Generative AI for text can be used to rewrite, summarize, personalize, translate, or extract data based on input texts. It can also be set up as a chatbot that end-users can interactively communicate with, and can be incorporated into other technologies like search engines.
- Generative AI can also create images or videos.
- These AI systems can be controlled using text-based prompts which provide task instructions and input data. For instance, you could prompt it with:
    - "Write three distinct headlines for the following news article: <article text>"
    - "Summarize the following text: <article text>"
    - "Translate the following text into English: <article text>"
    - "Explain <issue> in easy to understand language"
    - "Create an image/video showing <description of image>"

**Limitations**
- Accuracy: This technology does not always output text that is accurate.
- Attribution: This technology can't accurately include footnotes or citations for information sources it uses to create its responses.
- Biases: The outputs from this technology can be biased based on the data used to train the system, which typically reflects common societal biases (e.g. racial or gender).

**Trends**
- This technology is already accessible by more than 100 million people and access to it by all types of people will only continue to increase.